\documentclass{emulateapj}

\usepackage{natbib}
\usepackage{graphicx}

\bibpunct{(}{)}{;}{a}{}{,}

\begin{document}                          

\title{Initial Ionization of Compressible Turbulence}   

\author{Yuexing Li, Mordecai-Mark Mac Low}
\affil{Department of Astronomy, Columbia University, New York,
NY 10027, USA}
\affil{Department of Astrophysics, American Museum of Natural
History, New York, NY 10024, USA}
\and 
\author{Tom Abel}
\affil{Department of Astronomy and Astrophysics, The Pennsylvania State
  University, University Park, PA 16802, USA} 
\email{yxli@astro.columbia.edu, mordecai@amnh.org, tabel@astro.psu.edu} 

\begin{abstract}
We study the effects of the initial conditions of turbulent molecular
clouds on the ionization structure in newly formed H~{\sc ii} regions,
using three-dimensional, photon-conserving radiative transfer in a
pre-computed density field from three-dimensional compressible
turbulence.  Our results show that the initial density structure of
the gas cloud can play an important role in the resulting structure of
the H~{\sc ii} region. The propagation of the ionization fronts, the
shape of the resulting H~{\sc ii} region, and the total mass ionized
depend on the properties of the turbulent density field.  Cuts through
the ionized regions generally show ``butterfly'' shapes rather than
spherical ones, while emission measure maps are more spherical if the
turbulence is driven on scales small compared to the size of the H~{\sc ii}
region. The ionization structure can be described by an effective 
clumping factor $\zeta=\langle n \rangle \cdot \langle
n^2\rangle/\langle n\rangle^2$, where $n$ is number density of the gas. The
larger the value of $\zeta$, the less mass is ionized, and the more
irregular the H~{\sc ii} region shapes.  Because we do not follow
dynamics, our results apply only to the early stage of ionization when
the speed of the ionization fronts remains much larger than the sound
speed of the ionized gas, or Alfv\'en speed in magnetized clouds if it is
larger, so that the dynamical effects can be negligible.
\end{abstract}


\section{INTRODUCTION}
H~{\sc ii} regions are photoionized regions surrounding young OB stars. The
ionization structures and physical properties of these regions are
important to understanding of the formation of massive stars and their
feedback to the environment. Over the past several decades, H~{\sc ii}
regions have been extensively observed in various size and shapes
(e.g. see reviews by \citealt{wood89, garay99, churchwell99,
churchwell02}). Depending on their sizes, they are usually classified
as ultracompact (linear size below 0.1~pc), compact (0.1--1~pc), or
extended (several to tens of parsecs). Ultracompact and compact H~{\sc ii}
regions are ionized by massive stars still embedded in their
natal molecular clouds, while extended H~{\sc ii} regions are thought to be
in their mature states, powered by a combination of stellar radiation,
stellar wind and supernova explosions in the associated OB star
clusters (Yang et al. 1996). Despite the diversity of sizes, H~{\sc ii}
regions generally display common features such as inhomogeneities and
irregular shapes. One interesting and still open question on the
origin of these complex structures is: are they from the initial
conditions of the gas before ionization, or formed by dynamical
processes during H~{\sc ii} region evolution?

Early analytical and numerical work focused on the formation of H~{\sc
ii} regions and the propagation of ionization fronts (I-fronts), by
modeling the ionization of a massive star suddenly turned on in a
uniform gas (\citealt{stromgren39, kahn54, axford61, goldsworthy61,
tenorio-tagle76, elmegreen76}). It was soon realized, however, that
the interstellar medium is far from homogeneous. Simple condensations
or smooth variations of density were then included in some model
calculations (\citealt{flower69, marsh70, kirkpatrick70,
kirkpatrick72, pequignot78, koppen79, icke79, tenorio-tagle81}, see
\citealt{yorke86} for a review). However, in the early studies,
radiative transfer effects were either ignored in calculations or
limited to one or two-dimensional treatments, and suffered from low
resolutions (\citealt{sandford73, klein78, koppen78, tenorio-tagle79,
sandford82}), which limited their applications to observations.

Recently, several authors have reported two-dimensional simulations of
the dynamical evolution of H~{\sc ii} regions and their applications
to observations. \citet{garcia96} presented two-dimensional
gasdynamical simulations of the evolution of H~{\sc ii} regions in
constant and power-law density profiles, and argued that a thin-shell
instability of the ionization fronts could produce the irregular and
bright edges of the H~{\sc ii} region. \citet{williams99} also performed
two-dimensional gasdynamical simulations and showed that shadowing
instability could also lead to formation of dense clumps. \citet{freyer03}
presented two-dimensional radiation gasdynamical simulations of the
interaction between an isolated massive star and its homogeneous and
quiescent ambient ISM, and showed that the redistribution of mass by
the action of the stellar wind shell could form the ``finger-like''
shapes.

However, since star forming regions display a wide range of kinematic
properties and density distributions, the gas may well be turbulent, clumpy,
filamentary, and coupled with the magnetic field (\citealt{elm93, dyson95,
redman98, maclow99, crutcher99, williams00, khm00, osg01}; see \citealt{mk04}
for a recent review). These uniform models may not capture the realistic
density field of the ambient medium. The initial inhomogeneities and turbulent
structures of the gas clouds may affect the propagation of the ionization
fronts and contribute to the shaping of the resulting H~{\sc ii} regions. 

To test this, a reliable, three-dimensional treatment of radiative
processes is essential, and high resolution simulations of radiative
transfer fully coupled with gasdynamics in turbulent medium are highly
desirable. However, incorporating radiative transfer in gasdynamics
simulation presents a serious numerical challenge. The difficulties are due to 
the non-locality of the radiation physics, which hampers efficient
parallelization of the simulations and severely limits the size of problem
that can be tackled. So far most calculations have not been able to include
transfer of ionizing radiation into three-dimensional gasdynamics simulations,
so, as a step forward to investigate the ionization of turbulent clouds, we
apply radiative transfer to a static, pre-computed, three-dimensional,
turbulent density field. It is a simplified model, but so long as the speed of
the I-front is much larger than the sound speed of the ionized gas,
the expansion of the gas can be neglected \citep{spitzer78}, and the density
field can be treated as static. In cases where a magnetic I-front is
considered (e.g. \citealt{williams00}), the Alfv\'en speed in the mostly 
neutral gas ahead of the I-front should replace the ionized sound speed in the
above criterion if it is larger.  However, usually the ionized sound speed
dominates. 

As we will show, this simple approach produces a variety of
interesting morphological structures that may provide us some
understanding of the physical processes in H~{\sc ii} regions even
before more sophisticated calculations can be performed. However, 
our results only apply to the initial photoionization of static turbulence
before gasdynamics becomes important, so future comprehensive simulations
with a full treatment that couples radiative transfer and gasdynamics
will be necessary to confirm and extend them.

We present high resolution computations of the initial ionization of
compressible turbulence by a point ionizing source, such as a young
star. We apply INONE, a three-dimensional radiative transfer code
developed by \citet{abel00}, to a compressible turbulent medium from
different hydrodynamical (HD) and magnetohydrodynamical (MHD) models
using ZEUS-3D (\citealt{stone92a, stone92b}), as described in
\citet{maclow99}. INONE integrates the jump condition of I-fronts, and
uses a photon-conserving method that is independent of numerical
resolution and ensures correct propagation speeds for the I-fronts
\citep{abel99}. In \S\ref{sec_com} we briefly describe the methods of
computation, and the codes and models used in this work; we present
the results of propagation of I-fronts, emission measure maps, and
ionization structures in \S\ref{sec_result}; and in
\S\ref{sec_discussion} we give an analytical discussion of the key
parameters that determine the ionization structure, as well as the
validity of the assumptions.

\section{COMPUTATIONS}
\label{sec_com}
\subsection{Compressible Turbulent Molecular Clouds}
\label{subsec_sim}
Star forming molecular clouds have been observed to be highly clumpy and have 
broad emission line widths, which suggests supersonic motions in the 
clouds \citep{blitz93}. These supersonic motions seem not to be
ordered. \citet{crutcher99} presented Zeeman observations of magnetic
fields in molecular clouds that showed that velocities in the observed
clouds are typically Alfv\'enic. So molecular clouds are generally
self-gravitating, clumpy, magnetized, turbulent, compressible fluids. Although 
numerical simulations of transient, compressible turbulent molecular clouds
have been carried out for several years, we are far from drawing a
comprehensive picture of the complicated structure of the clouds.

\citet{maclow99} conducted direct numerical computations of uniform,
randomly driven turbulence with $128^3$ resolution using the ZEUS-3D MHD code ,
which successfully simulated the density distribution of the turbulent
medium. ZEUS-3D is a well-tested, Eulerian, finite-difference code
\citep{stone92a, stone92b, clarke94}. It uses second-order \citet{vanleer77}
advection and resolves shocks using von Neumann artificial viscosity. It also
includes magnetic fields in the MHD approximation
\citep{hawley95}. \citet{falle02} pointed out some problems with ZEUS:
rarefaction waves often break up into a series of jumps, and adiabatic
MHD shocks sometimes show errors.  However, as we are using isothermal
MHD, and are primarily concerned with the density field, which is
barely affected by rarefaction shocks, our models remain valid. 

\begin{deluxetable}{cccc}
\label{tab1}
\tablecolumns{4}
\tablewidth{3.0in}
\tablecaption{Turbulence Models \citep{maclow99}}
\tablehead{\colhead{$Model$} & \colhead{$\dot{E_{\rm in}}$\tablenotemark{a}} &
\colhead{$k$\tablenotemark{b}} & \colhead{$C$\tablenotemark{c}}}
\startdata
HA8  & 0.1 & 8 &  1.50  \\
HC2  & 1   & 2 &  5.88  \\
HC4  & 1   & 4 &  4.89  \\
HC8  & 1   & 8 &  2.97  \\
HE2  & 10  & 2 &  7.24  \\
MA81 & 0.1 & 8 &  1.46  \\
MC41 & 1   & 4 &  2.46  \\
MC45 & 1   & 4 &  4.35  \\
MC4X & 1   & 4 &  5.88  \\
MC81 & 1   & 8 &  2.46  \\
\enddata 
\tablenotetext{a}{Energy Input}
\tablenotetext{b}{Driving Wavenumber}
\tablenotetext{c}{Clumping Factor}
\end{deluxetable}

For our computations, we assemble $384^3$ density fields by repeating
the periodic $128^3$ models of \citet{maclow99}. Table 1 describes the
models we use. The model names begin with either ``H" for hydrodynamic
or ``M" for MHD, then have a letter from ``A" to ``E" specifying the
level of energy input $\dot{E_{\rm in}}$, then a number giving the
dimensionless wave-number $k$ chosen for driving, and then, for the
MHD models, another number indicating the initial field strength
specified by the ratio of Alfv\'en speed to the sound speed. In the
last column, $C$ is the gas clumping factor defined as $C=\langle
n^2\rangle/\langle n\rangle^2$. Note that the values of $C$ listed here
are averaged over the whole simulated volume.

\subsection{Ionization Front Tracking}
\label{subsec_rt}
In general, a study of the evolution of ionization zones around a ionizing
source requires a full solution to the radiative transfer equation
\citep{kirchhoff60}: 
\begin{equation}
\label{rt_kir}
\frac{1}{c}\frac{\partial I_{\rm \nu}}{\partial t} + \hat{n} \cdot \nabla
I_{\rm \nu} = \eta_{\rm \nu} - \chi_{\rm \nu}I_{\rm \nu} \,,
\end{equation} 
where $\hat{n}$ is a unit vector along the direction of the radiation, $I_{\rm
\nu}$ is the monochromatic specific intensity of the radiation field, 
and $\eta_{\rm \nu}$ and $\chi_{\rm \nu}$ are emission and absorption
coefficients, respectively. 

However, a direct solution of equation (\ref{rt_kir}) is usually
impractical because of its high dimensionality. \citet{abel99} made
the calculations feasible by developing a ray tracing algorithm for
radial radiative transfer around point sources, reducing the
dimensionality of the transfer equation to a level where ionization
can be computed on a Cartesian grid. This algorithm conserves
energy explicitly and thus gives the right speed of I-fronts, but
ignores the diffuse field produced by scattered radiation.

\citet{kahn54} defined the nomenclature of $R$ (``rarefied'') and $D$
(``dense'') fronts according to the speed of the I-fronts with respect
to the sound speed of the ionized gas: R-type fronts move
supersonically while D-type fronts move subsonically. If a source of
ionizing radiation suddenly switches on, the I-front is initially weak
R-type. Once its velocity drops to about twice the sound speed in the
ionized gas, the I-front becomes D-type \citep{dyson02}. The
propagation of an R-type I-front in a static medium is given as:
\begin{equation}
\label{eq_if}
4\pi R_{\rm I}^2n_{\rm {H}}(R_{\rm I})\frac{dR_{\rm I}}{dt}=F-\int_0^{R_{\rm
I}}\alpha_{\rm B} 4\pi r^2 n_{\rm p}(r)n_{\rm e}(r)\ dr\,.
\end{equation}
where  $R_{\rm I}$ is the radius of the I-front, $F$ is the ionizing photon
flux, $n_{\rm H}$, $n_{\rm p}$ and $n_{\rm e}$ are the number density of the
neutrals, ions, and electrons, respectively. We assume a constant case B
recombination rate, $\alpha_{\rm B} = 2\times 10^{-13}$ cm$^3$ s$^{-1}$ for $T
\sim 10^4$ K. Integrating equation (\ref{eq_if}) along the rays with the
ray-tracing technique mentioned above gives the time at which the 
ionization front arrives at a given cell, as is done in INONE
\citep{abel00}. Retrieving the arrival time in a 3D array allows one to
investigate the time dependent morphology of the I-fronts.

\subsection{Scaling}
\label{subsec_scaling}
In this paper, we use INONE to compute photoionization of different
turbulence models. For simplicity, we consider hydrogen gas only. We expect
the inclusion of helium to have only a minor effect on the results presented
here. The 
simulations are scale free and depend only on the ionizing flux $F$
and density $n$ of the medium, so one can derive the scaling of
physical parameters such as ionized mass $M$, ionized volume $V$, and
ionization time $t$ in terms of $F$ and $n$. As the analytical derivation in
\S\ref{sec_is} demonstrates, the ionized mass at late times scales as $M
\propto F/n$; the ionized volume at late times is close to the volume of the
Str\"omgren sphere, and scales as $V \propto F/n^2$; and the characteristic
timescale is the recombination time, which scales as $t \propto 1/n$. As an
example, if we choose $F_0 = 1.2 \times 10^{49}$ s$^{-1}$, which is typical for
an O6 star \citep{panagia73}, and a cloud with a size of 0.5 pc and average
density of $n_0 = 5\times 10^3$ cm$^{-3}$, we then have $M_0 = 10 M_{\odot}$,
$V_0 = 0.08$ pc$^{3}$, and $t_0 = 10^{11}$ s (these will be given in
\S\ref{sec_result} as the results of the simulations). For given F and n, one
can rescale our results as follows:
\begin{eqnarray}
\label{eq_mas}
M_0' & = & \left(\frac{F}{F_0}\right)\left(\frac{n}{n_0}\right)^{-1}M_0 \nonumber \\
     & = & (10 \ M_{\odot})\left(\frac{F}{1.2\times 10^{49}\
     s^{-1}}\right)\left(\frac{n}{5\times  10^3 \ cm^{-3}}\right)^{-1}\,,
\end{eqnarray}
\begin{eqnarray}
\label{eq_vol}
V_0' & = & \left(\frac{F}{F_0}\right)\left(\frac{n}{n_0}\right)^{-2}V_0  \nonumber \\
     & = & (0.08 \ pc^{3}) \left(\frac{F}{1.2\times 10^{49}\
     s^{-1}}\right)\left(\frac{n}{5\times  10^3\ cm^{-3}}\right)^{-2}\,,
\end{eqnarray}
\begin{eqnarray}
\label{eq_tim}
t_0' & = & \left(\frac{n}{n_0}\right)^{-1}t_0 \nonumber \\
     & = & (10^{11}\ s) \left(\frac{n}{5\times 10^3\ cm^{-3}}\right)^{-1}\,. 
\end{eqnarray}

In the simulations, we use the pre-computed gas density distribution with
ZEUS-3D for the HD and MHD models listed in Table 1 as input into INONE. We
will discuss the validity of this treatment later in \S\ref{subsec_valid}.
Three simulations are conducted for each model, by varying the position of the 
star to the maximum density, minimum density, or center of the density field
(random density), so for our example a total of 30 simulations were performed
on SGI Origin2000 computers. Each computation took several days on a single
processor.

\section{RESULTS}
\label{sec_result}

\subsection{Shapes of the Ionized Regions}
\label{subsec_butterfly}
\begin{figure}[h]
\epsscale{1.0}\plotone{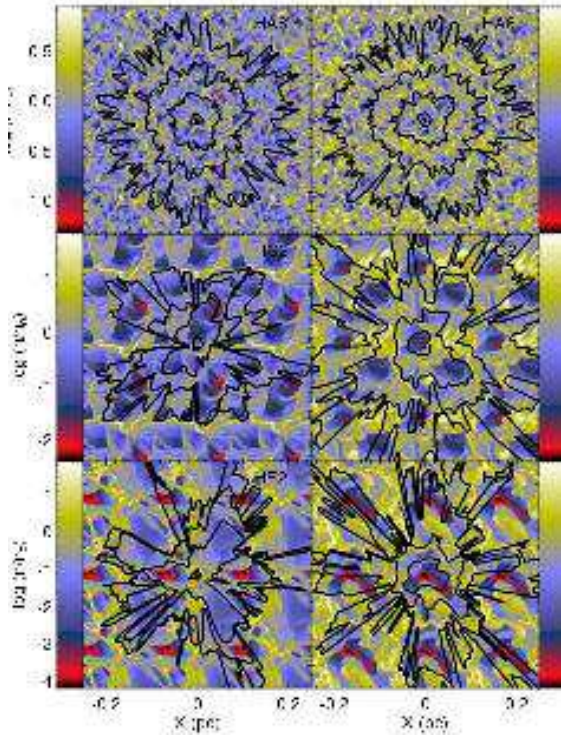}
\caption{\label{fig_hd} Propagation of an R-type I-front in a $384^3$ density
field simulated with different HD models. The ionizing source is
located at maximum (\textit{left}), and minimum density
(\textit{right}), respectively. The contours give the position of
the I-front from 0.1 yr to 100 yr, with the interval increasing evenly by a
factor of 10. The size of box is 0.5 pc, $\langle n_0 \rangle = 5\times 10^3$ cm$^{-3}$. Note
the image and contour are slices through middle point of the cube.} 
\end{figure}

\begin{figure}[h]
\epsscale{1.0}\plotone{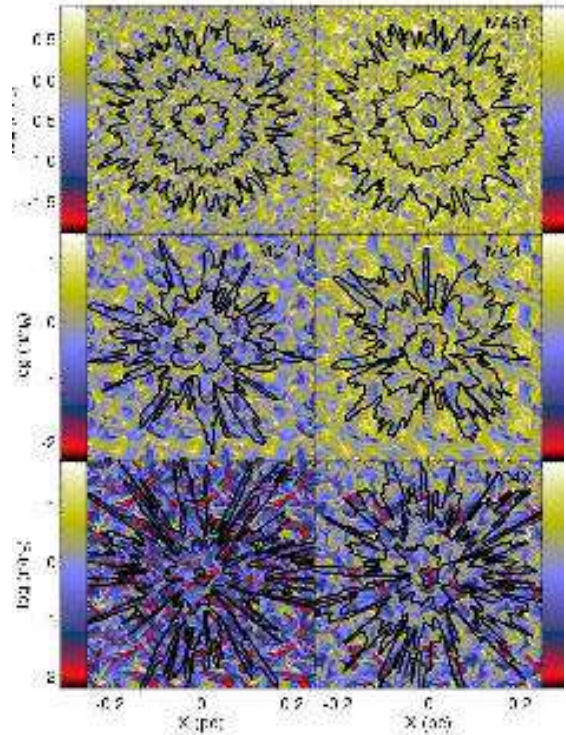}
\caption{\label{fig_mhd} Propagation of the I-front in a $384^3$ density
field simulated with different MHD models. The ionizing source is
located at maximum (\textit{left}), and minimum density
(\textit{right}), respectively. The contours give the position of 
the I-front from 0.1 to 100 years, with the interval increasing evenly by a
factor of 10. The size of box is 0.5 pc, $\langle n_0 \rangle = 5\times 10^3$
cm$^{-3}$. Note the image and contour are slices through middle point of the cube.}  
\end{figure}

Figures~\ref{fig_hd} and~\ref{fig_mhd} show the propagation of the
I-fronts into the turbulent medium surrounding a point source. The background
image is the density distribution of the gas simulated by HD and MHD models,
respectively. The contours are the arrival times of the I-fronts. In Figure
\ref{fig_hd}, the density fields are taken from HD models with different
energy input and driving wave number, while in Figure~\ref{fig_mhd},
the density fields are from MHD models with different driving wave
number and magnetic field strength, as listed in Table 1. In both
figures, the left panels are for the case that the point source (star) is
placed at the maximum density, while in the right panels the star is placed at
the minimum density. In each individual panel, the contours give the position
of the I-fronts in our example scaling from 0.1 to 100 years, with the
interval increasing evenly by a factor of 10. The models chosen in each figure
have different average clumping factors (see Table 1). 

At early times, the recombination term in equation (\ref{eq_if}) can be
neglected, and the I-front velocity is determined by the local density
structure of the gas: the front travels faster into voids and slower into
denser filaments. The shape of the I-front depends on the size and contrast of
the voids and clumps. Longer turbulent driving wavelengths (smaller driving
wave numbers) create structure in the density field on larger scales and
produce more asymmetric H~{\sc ii} regions with characteristic 'butterfly'
shapes. At later times, recombination becomes significant, increasing the
dependence on local density and making the resulting asymmetry more pronounced.

The initial I-front velocity is sensitive to the placement of the ionizing
point source, as can be seen from a comparison of the innermost contours in the
left-hand and right-hand columns of Figures~\ref{fig_hd}
and~\ref{fig_mhd}. However, this effect becomes less important at later times, 
and the final H~{\sc ii} regions of the same model have very similar sizes,
regardless the position of the ionizing source, as can be seen in
Figures~\ref{fig_vol}. 

In Figures~\ref{fig_hd} and~\ref{fig_mhd}, the images are sorted in order of
increasing clumping factor $C$; e.g. in Figure \ref{fig_hd} $C(\rm HA8) =
1.5$, $C(\rm HC2) = 5.88$, $C(\rm HE2) = 7.24$, and similar trend in
Figure~\ref{fig_mhd}. Comparison of the images clearly shows a strong link
between the size of the clumping factor and the morphology of the H~{\sc ii}
region -- the larger the clumping factor, the more asymmetric the H~{\sc ii}
region becomes. 

\subsection{Emission Measure}
For comparison with observations, we can map the emission measure in
our models,
\begin{equation}
\label{eq_em}
EM = \int n_{\rm e}^2 dl
\end{equation}
where $l$ is the size of the region along the line-of-sight. Typically more
compact H~{\sc ii} regions have higher emission measures.  For example, an
extended H~{\sc ii} region with size of 10--100 pc and density of 10 cm$^{-3}$ has 
EM in the range of 10$^{3-4}$ pc cm$^{-6}$, while an ultracompact
H~{\sc ii} region with size $< 0.1$~pc and density $\ge 10^4$
cm$^{-3}$ is brighter, with EM $> 10^7$ pc cm$^{-6}$.
\begin{figure}[h]
\epsscale{1.0}\plotone{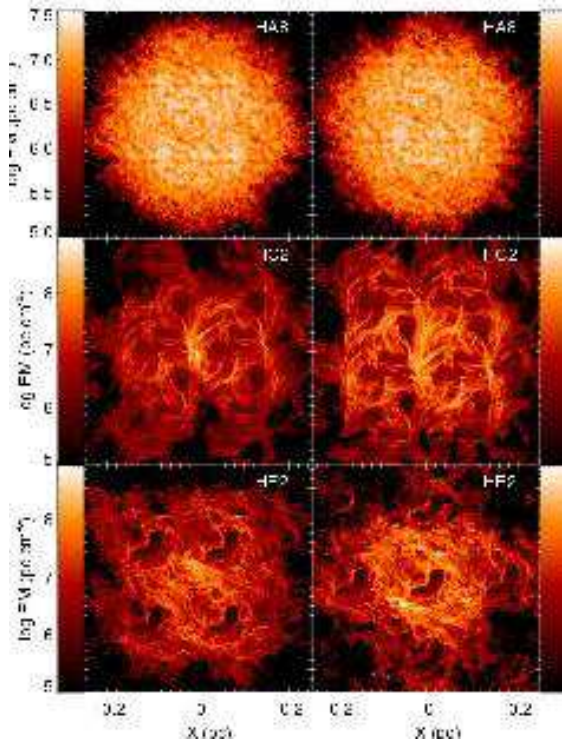}
\caption{\label{fig_em_hd}Emission measure of the ionized regions of
HD models at $t \simeq 100$ yr, assuming the scales used in the text
as an example. The ionizing source is located at maximum
(\textit{left}), and minimum density (\textit{right}), respectively.}
\end{figure}

\begin{figure}[h]
\epsscale{1.0}\plotone{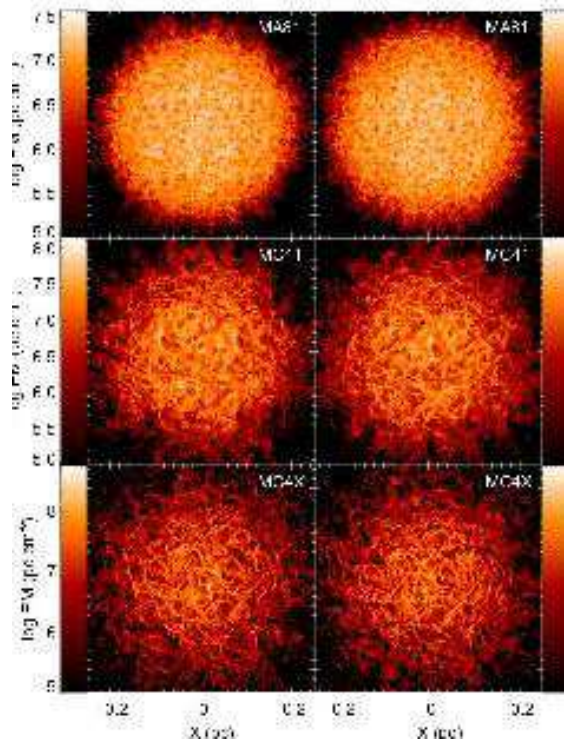}
\caption{\label{fig_em_mhd}Emission measure of the ionized regions of MHD
models at $t \simeq 100$ yr assuming the scales used in the text
as an example. The ionizing source is located at maximum
(\textit{left}), and minimum density (\textit{right}), respectively.}  
\end{figure}

We calculate emission measure maps from the simulations by integrating
density $n_{H}^2$ of the ionized regions through the path length in
the third direction. Figures
\ref{fig_em_hd} and~\ref{fig_em_mhd} show emission measure maps in the
x-y plane of HD and MHD models, respectively. The arrangement of the
models in both figures are the same as in Figures~\ref{fig_hd} and
\ref{fig_mhd}. We can see that the range of the EM is around $10^{6 -
8}$ pc cm$^{-6}$ for our adopted parameters, which agrees, as
expected, with the observed range of compact H~{\sc ii} regions. 

The shape and structure of the EM map depends on the clumping factor:
the smaller the clumping factor is, the more spherical the map
appears. Filamentary structures and contrast within the H~{\sc ii}
region also increase with $C$. There are some differences between the
HD and MHD models with $ C > 1$: HD models tend to have more irregular
shapes, and bigger voids and clumps.  Larger driving wavelengths of
the turbulence also produce larger voids and clumps. These results
generally suggest that the turbulent structures observed in H~{\sc ii}
regions may be driven at large scale, and that the initial gas clouds may have
a large clumping factor.  However, the influence of dynamics will have
to be computed to put any such conclusion on a firm basis.

\subsection{Ionization Structure}
\label{subsec_ion}
\begin{figure}[h]
\epsscale{1.0}\plotone{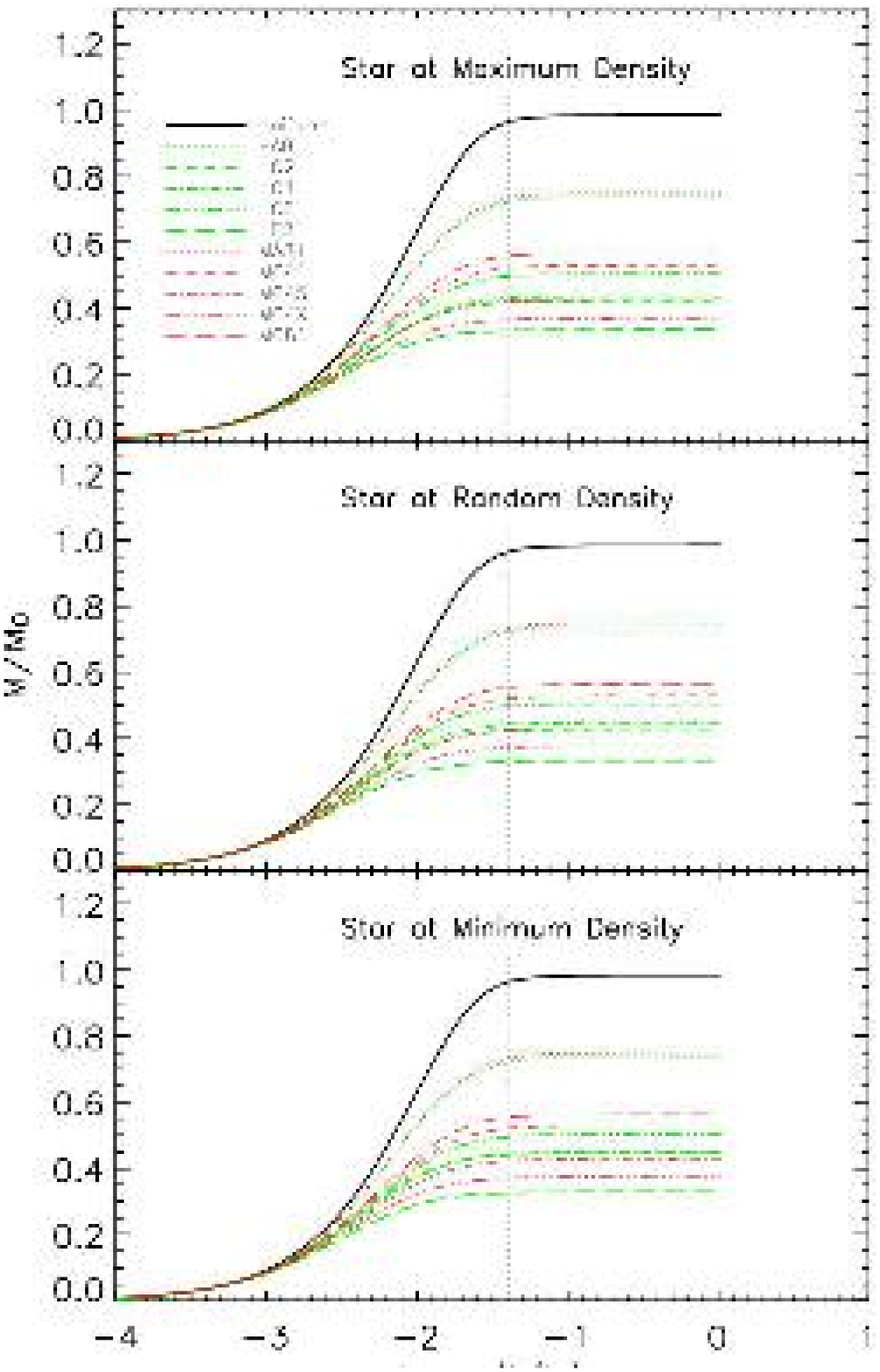}
\caption{\label{fig_imas}Ionized mass over time for different models, with ionizing star
at (top:) maximum density, (middle:) a random position, and (bottom:) minimum
density. $M_0$ and $t_0$ are defined in \S\ref{subsec_scaling}. The vertical
dashed line indicates the minimum time when $v_{\rm if} = 2c_s$ (see Figure
\ref{fig_vel}).} 
\end{figure}
\begin{figure}[h]
\epsscale{1.0}\plotone{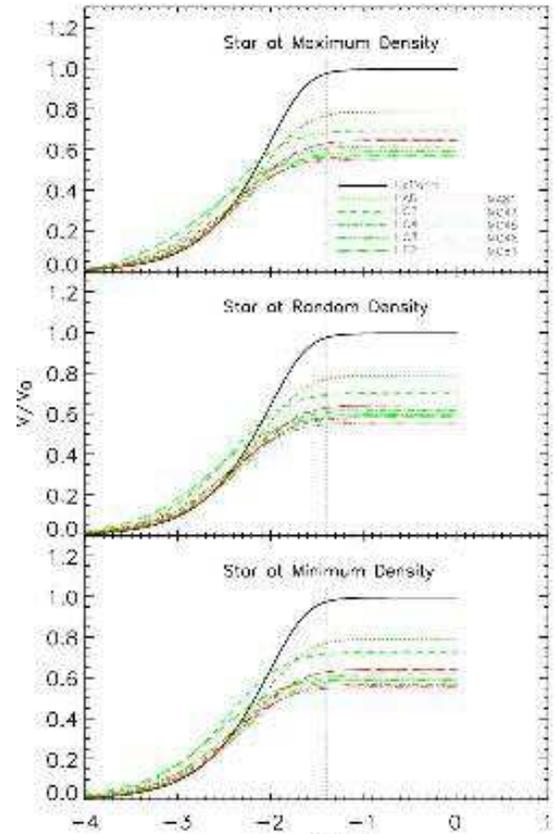}
\caption{\label{fig_vol}Ionized volume over time for different models, with ionizing star
at  (top:) maximum density, (middle:) a random position, and (bottom:) minimum
density. $V_0$ and $t_0$ are defined in \S\ref{subsec_scaling}. The vertical
dashed line indicates the minimum time when $v_{\rm if} = 2c_s$ (see Figure
\ref{fig_vel}).}  
\end{figure}

Figures~\ref{fig_imas} and~\ref{fig_vol} show the time-dependent
ionized mass and ionized volume, respectively. Note that the units, $M_0$, $V_0$,
and $t_0$ are the same as in equations (\ref{eq_mas}), (\ref{eq_vol}), and
(\ref{eq_tim}). Different panels have the ionizing star at different positions,
while within each panel, different curves indicate different models. The ionized
volume is calculated by integrating the ionized cells at time t, and the ionized mass
is the integral of the mass encompassed in the ionized volume. The curves
become saturated around $t \sim 100$ years. It appears
that the mass of the ionized gas in our example is close to that of a compact
H~{\sc ii} region \citep{franco00}. For the same input energy strength and driving
wave number, no big difference is seen between the MHD and the HD
models. However, all the curves follow a sorted
order of the clumping factor $C$, with HA8 and MA81 having the smallest $C$,
while HE2 has the largest $C$. 

\subsection{Compared With Two-Phase Clumpy Model}
\label{subsec_2phase}
Here we compare our results with an analytic model of a two-phase clumpy
medium. Consider a  medium with optically thin clumps of number
density $n_{\rm cl}$ in a uniform density field of number density $n_{\rm
bg}$, which was studied by \citet{koppen79}. If we take a filling factor $\epsilon$ for
clumps, and the density contrast $\xi=n_{\rm bg}/n_{\rm cl}$, then the average
density is: 
\begin{equation}
\bar{n}=\epsilon n_{\rm cl} + (1-\epsilon)n_{\rm bg} = n_{\rm
cl}\left[\epsilon + (1-\epsilon)\xi \right]\,,
\end{equation}
The ionization structure of this medium is:
\begin{eqnarray}
\label{eq_clumpy}
4\pi R_{\rm I}^2\bar{n}\frac{dR_{\rm I}}{dt} & = & F - \int_0^{R_{\rm
I}}4\pi r^2\alpha_{\rm B} n_{\rm cl}^2\epsilon dr \nonumber \\
&  & - \int_0^{R_{\rm I}}4\pi r^2\alpha_{\rm B} n_{\rm bg}^2(1-\epsilon)dr \,,
\end{eqnarray}
where $\alpha_{\rm B}$ is the recombination-rate coefficient as in equation
(\ref{eq_if}). Define a two-phase clumping factor $C_{2}$,
\begin{equation}
\label{eq_c2_1}
C_2 = \frac{\epsilon+(1-\epsilon)\xi^2}{\epsilon\xi+(1-\epsilon)\xi^2} \,,
\end{equation}
and define $dN=4\pi R_{\rm I}^2\bar{n}dR_{\rm I}$, where $N$ is the total
number of ions. Equation (\ref{eq_clumpy}) can then be written as
\begin{equation}
\label{eq_c2_2}
\frac{dN}{dt}=F - \alpha_{\rm B} C_2 n_{\rm bg}N \,,
\end{equation}
from which we get an analytical solution of the number of the ions N, and
the ionized mass $M_{\rm {2}}$:
\begin{equation}
\label{eq_imas}
M_{2} = m_p \cdot N=\frac{F}{\alpha_{\rm B} C_2 n_{\rm bg}}(1-e^{-\alpha_{\rm B} C_2 n_{\rm bg}t})\,.
\end{equation}
\begin{figure}[h]
\epsscale{1.1}\plottwo{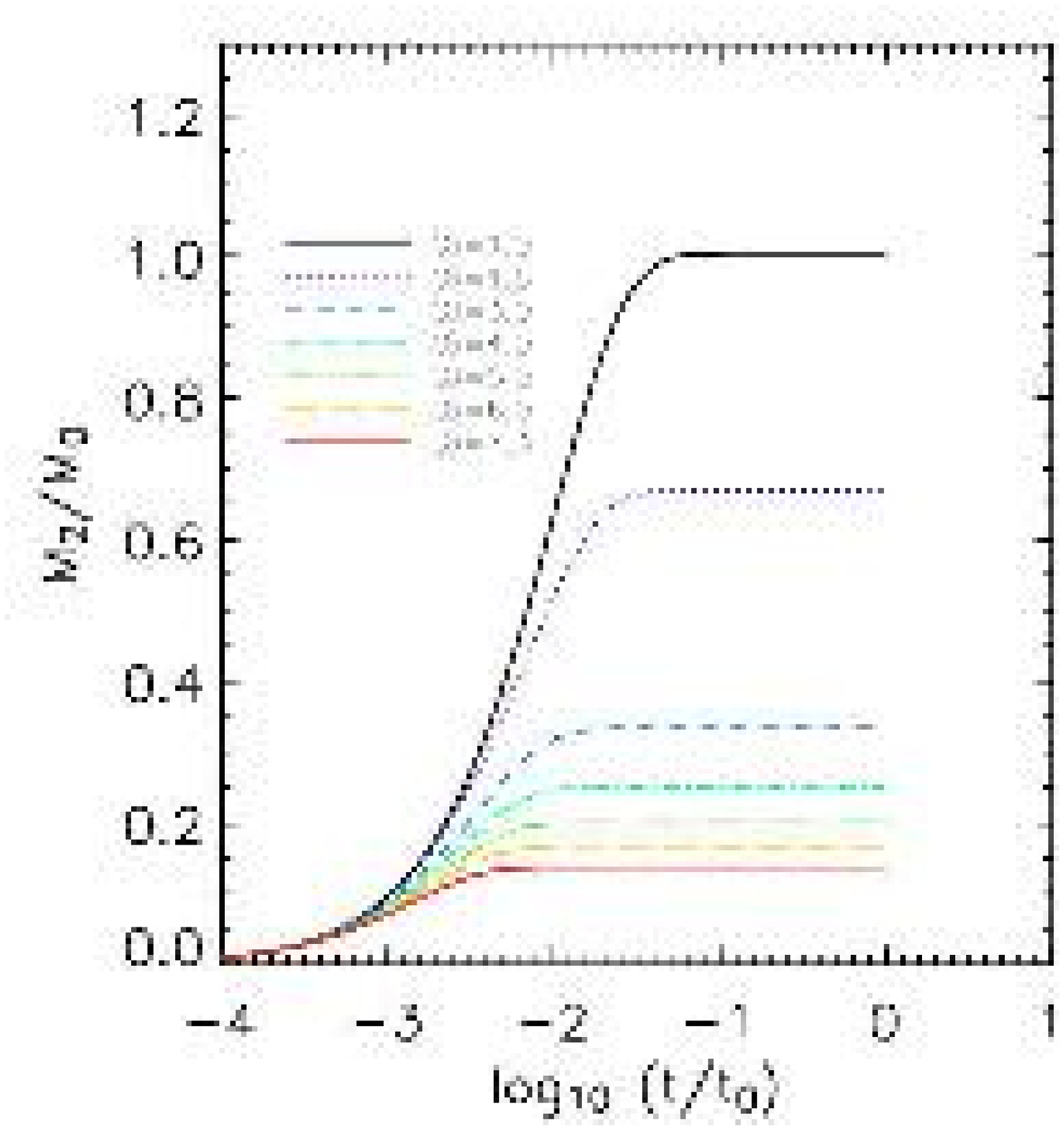}{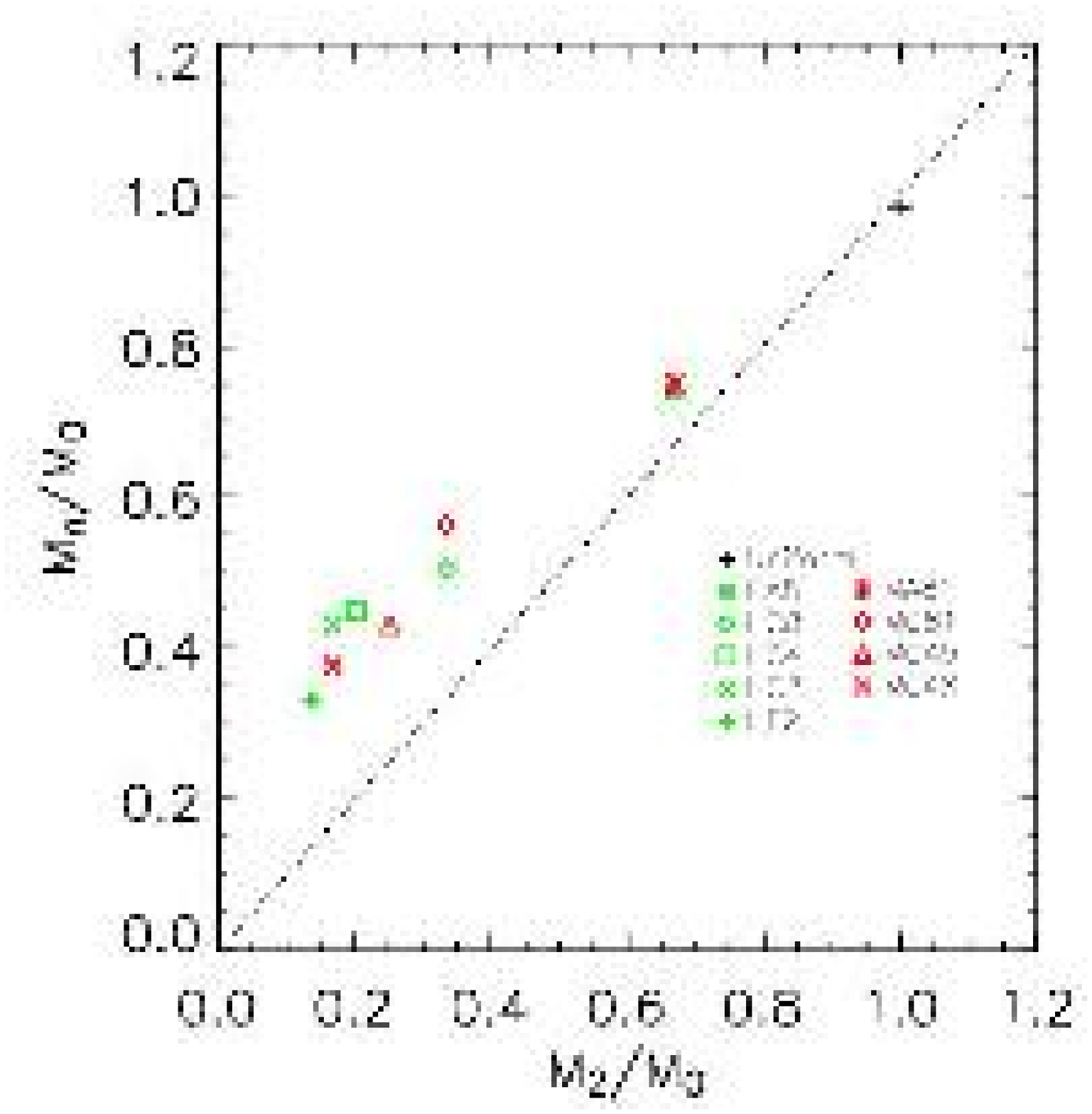}
\caption{\label{fig_2phase}\textit{Left}: ionized mass of two-phase clumpy model over
time for different two-phase clumping factors $C_2$ taken after the values of the
turbulence models in Table 1. \textit{Right}: comparison of numerical ionized mass $M_n$
(the ionizing star is located at random position) to analytical ionized mass
from the two-phase models $M_{\rm {2}}$. Note that the dotted line in this plot is
not a fitting curve, but just a diagonal line to guide the eyes.} 
\end{figure}
Figure~\ref{fig_2phase} shows the ionized mass from the two-phase model, and a
comparison between the analytical results and the numerical ones. The clumping
factors $C_2$ were chosen to be close to the values measured in the turbulence
simulations, as listed in table 1. From top to bottom in the plot, we have 1.0
(uniform gas), 1.5, 3.0, 4.0, 5.0, 6.0 and 7.3 respectively. The background
density $n_{\rm bg}$ is taken as the mean density $n = 5\times10^3$ cm$^{-3}$
of the scaled simulations. Though similar in shape, the results of the
two-phase model are distinct from those of the turbulent models; the ionized
mass for a given $C_2$ is less than in the corresponding turbulent model, and
the difference increases as the medium becomes more clumpy. 

\section{DISCUSSION AND SUMMARY}
\label{sec_discussion}
We have presented simulations of different models, and found that a two-phase
clumpy model can not fully describe the ionization structure of the
turbulence models. However, some questions still remain:
what determines the ionization structure? When are our assumptions valid?

\subsection{What Determines the Ionization Structure?}
\label{sec_is}
The results shown above suggest that the ionization structure is sensitive to
the input energy strength and driving wave number of the turbulence. But
qualitatively, what is the key factor that determines the ionization structure?  

Assume the density field is constant, and the gas is fully ionized within the
ionization front, then both the ion density $n_{\rm p}(r)$ and the electron 
density $n_{\rm e}(r)$ at radius $r$ are equal to the original hydrogen density
$n(r)$, $n_{\rm p}(r) = n_{\rm e}(r) = n(r)$. Let $N(t)$ be the total number
of the ions at time t, $N(t) = \int_0^{R_{\rm I}}n(r)\ dV$, then equation
(\ref{eq_if}) can be rewritten as: 
\begin{eqnarray}
\label{dndt1}
\frac{dN(t)}{dt} & = & F-\alpha_{\rm B}\int_0^{R_{\rm I}} n(r)^2 \ dV  \nonumber \\
		 & = & F-\alpha_{\rm B} \cdot\langle n\rangle \cdot \frac{\langle n^2\rangle}{\langle
n\rangle^2} \cdot N(t) \,. 
\end{eqnarray}

We can now see that the ionized mass depends explicitly on the average gas density
$\langle n\rangle$, and the local clumping factor ($C=\langle
n_H^2\rangle/\langle n_H\rangle^2$) within the I-front. Different driving
strength and wave number of the turbulence yield different size and contrast
of clumps and voids, which is represented by $C$ and $\langle n\rangle$, thus yielding
different ionization structures. Since $R_{\rm I}$ changes with time, we
further define an effective clumping factor $\zeta(t) = \langle n \rangle
\cdot \langle n^2\rangle/\langle n\rangle^2$, so that equation (\ref{dndt1})
can be rewritten as:    
\begin{equation}
\label{if_ana1}
\frac{dN(t)}{dt}= F-\alpha_{\rm B} \cdot \zeta(t) \cdot N(t) \,.
\end{equation}

\begin{figure}
\epsscale{1.0}\plotone{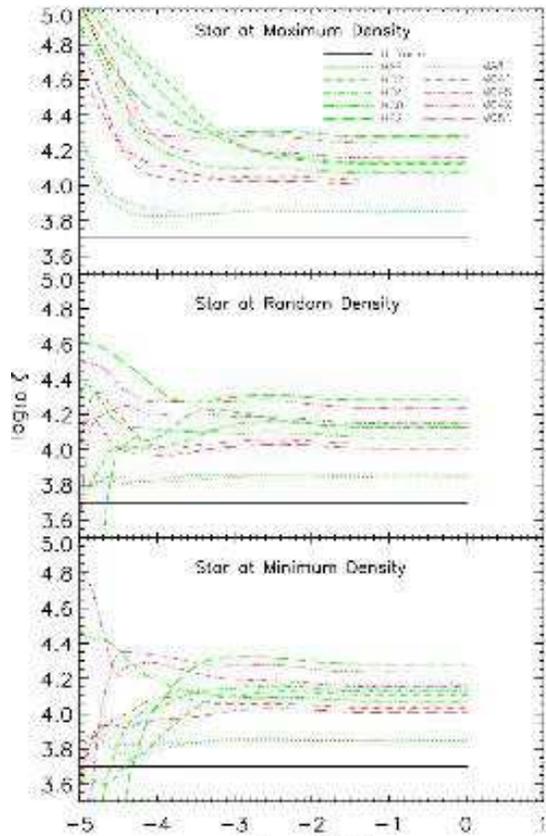} 
\caption{\label{fig_clum}Effective clumping factor $\zeta$ versus time as an indicator of
ionization structure. It varies with different models, and different
position of the ionizing star at (top:) maximum density, (middle:) a random
position, and (bottom:) minimum density.} 
\end{figure}

From this we can see that $\zeta$ describes the density field, and determines the
ionized mass. Figure~\ref{fig_clum} shows  $\zeta$ as a function of
time for our models. Comparing Figure~\ref{fig_clum} with Figure
\ref{fig_imas}, we find that the order of the models in Figure~\ref{fig_clum}
is exactly the inverse order in Figure~\ref{fig_imas}. The larger the
$\zeta$, the less mass is ionized, as predicted from equation
\ref{if_ana1}. 

The dependence of the effective clumping factor $\zeta(t)$ on the density
field is nontrivial because the averaging volume is itself a time-dependent
function of the density field. However, $\zeta$
tends to saturate once the radius of the I-front is larger than the turbulence
driving scale. This also corresponds to the saturation of the
photoionization. The total ionized mass can then be approximated well with
equation~\ref{if_ana1}. Note also that in these figures, different positions
of the ionizing star does not make much difference, because $\zeta$ is a
measure of the density field, and the ionized mass depends only on $\zeta$ and
the ionizing flux, no matter where the source is. In turbulence simulations,
resolution might affect the value of $\zeta$, but it would not affect our
result as a function of $\zeta$. 
\begin{figure}
\epsscale{1.0}\plotone{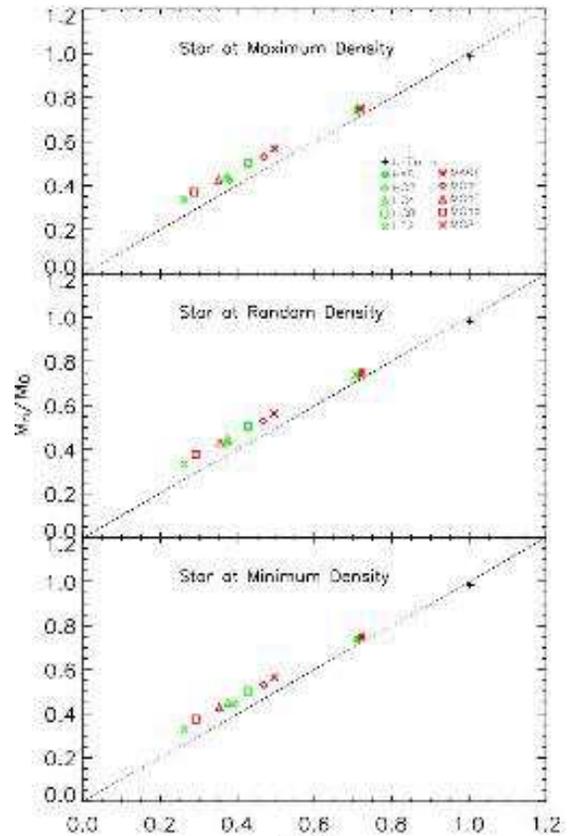}
\caption{\label{fig_mas_comp}Comparison of numerical ionized mass $M_n$ to
analytical ionized mass $M_a$ for different models, and different position of the ionizing
star at (top:) maximum density, (middle:) a random
position, and (bottom:) minimum density. Note that the dotted lines in the plots are
not fitting curves, but just diagonal lines to guide the eyes.} 
\end{figure}

Figure~\ref{fig_mas_comp} shows a comparison of numerical and
analytical values of ionized mass for the different models. The small variance
between numerical simulations and analytical calculations 
suggests that the calculations are self-consistent, and that
the numerical resolution of the calculation is good enough to resolve the
structure. Since in reality the density fields are more complicated than the
simplified model used in the analytical calculations, there is a systematic
difference between the numerical results and the analytical ones: in all these
models, the smaller the clumping factor, the smaller the difference.   

\subsection{When Are the Assumptions Valid?}
\label{subsec_valid}

\begin{figure}
\epsscale{1.0}\plotone{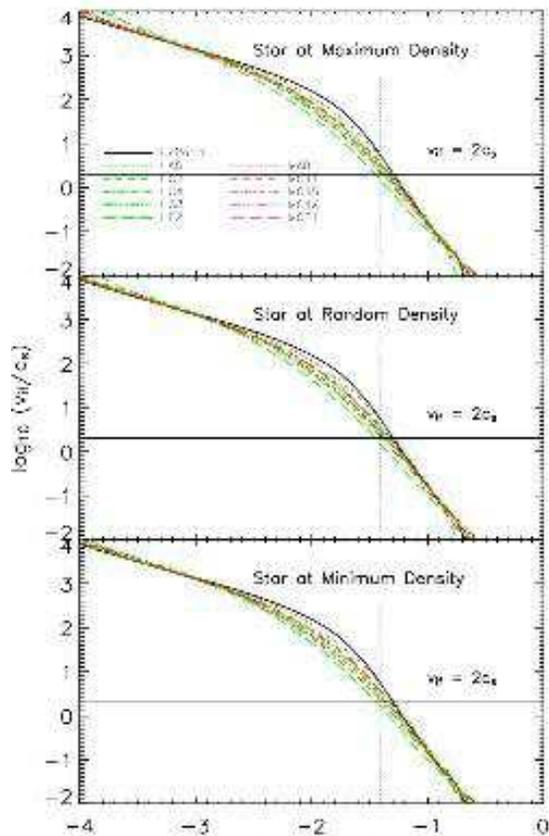}
\caption{\label{fig_vel}Time evolution of $v_{\rm if}/c_s$, ratio of speed of
I-fronts $v_{\rm if}$ to sound speed $c_s$, for different models. When $v_{\rm if}$
decreases to twice of the sound speed, expansion of the ionized gas is
significant. The vertical dashed line indicates the minimum time at when
$v_{\rm if} = 2c_s$.} 
\end{figure}

In the above simulations, the density field of the gas is simulated beforehand
and then input to INONE, so gasdynamics is not computed simultaneously
during the course of radiative transfer. However, when the speed of
the I-fronts slows to roughly twice the sound speed of the ionized gas, the
expansion of the gas will dominate the evolution of the region, and gasdynamics
must be considered \citep{spitzer78}. Figure~\ref{fig_vel} shows the time
evolution of the ratio of speed of I-fronts $v_{\rm if}$ to sound speed $c_s$,
$v_{\rm if}/c_s$, for different models. The sound speed is $c_s = 9.1$
km s$^{-1}$ for $T = 10^4$ K. The velocity of 
the I-fronts, $v_{\rm if} = dR_i/dt$, is hard to derive because it is three
dimensional, and even at the same photon arrival time, it may vary
dramatically in different cells if the local clumping factors are different,  so
what is shown here is an average value at $R_i$, the radius of the 
I-front at time t. However, one should keep in mind that in some regions
$v_{\rm if}$ may be much smaller than the average value. For example, an
obliquely propagating ionization front moves slowly and transitions from R-type to
D-type more quickly (\citealt{williams99}). Fully dynamical simulations are necessary 
to assess the impact of these effects. From Figure~\ref{fig_vel}, we can see
that the speed of I-fronts decreases rapidly 
with time, and at about log$_{10}(t/t_0) = -1.4$ (which corresponds to $t \sim
100$ years in our example), it drops to twice of the sound speed, so after
this time the simulations are likely unreliable. From Figure~\ref{fig_imas}
and~\ref{fig_vol}, we can see that this is the time when the curves of ionized
mass and ionized volume become saturated, so the details of the ionization
structure derived up to that point are likely reliable. 

\citet{williams00} presented one-dimensional simulations on jump
conditions of magnetic I-fronts and showed that fast R-type I-fronts
could decrease quickly to slow D-type I-fronts in oblique magnetic
fields. We note, however, that as long as the propagation speed of the
I-fronts is much larger than the Alfv\'en speed of the medium,
this effect remains unimportant. In molecular clouds, the typical
Alfv\'en speed is about 1--5~km~s$^{-1}$, which is no larger than the
sound speed in ionized gas.  Table 1 shows that the ratio of Alfv\'en
speed to the sound speed in the MHD models is appropriate for cold,
molecular clouds. \citet{crutcher99} showed an empirical
power-law relationship between magnetic field strength and density
above $10^{3-4}$~ cm$^{-3}$. According to this scaling law, our
density field ($\langle n \rangle = 5000$ cm$^{-3}$) might have $B \sim
50 \ \mu G$, which gives an Alfv\'en velocity $v_A =
(B^2/(4\pi\rho))^{1/2} \sim 5$ km s$^{-1}$. 
The ionized sound speed $c_s = 9.1$ km s$^{-1}$, larger than the Alfv\'en
speed $v_A$, which suggests that our critical point log$_{10}(t/t_0) =
-1.4$ occurs before the magnetic field becomes important. 
 
Again, it should be emphasized that our simulations are simplified
models in which a point ionizing source is turned on in a static
turbulent density field. We did not take into account several effects,
such as the mass accretion by the ionizing star, the diffuse field,
and any time evolution in the ionizing luminosity of the star
\citep{yorke86}. These effects should eventually be included in more
comprehensive future simulations. Our main focus here is to
investigate how the initial conditions of the gas affect the
ionization structures of the resulting H~{\sc ii} region.

\subsection{Summary}
We have presented a study of the effects of turbulent initial
conditions on the ionization structures of young H~{\sc ii}
regions. We performed high resolution, three-dimensional radiative
transfer computations, as well as analytical calculations, of the
propagation of I-fronts into static density distributions drawn from
numerical simulations of compressible, turbulent molecular clouds.
Our results show that the initial turbulent structure of the gas can
play an important role in the early ionization structure. The
propagation of the I-fronts, the shape of the resulting H~{\sc ii}
regions, and the total mass ionized depend on the local strength of
the voids and clumps.  We can characterize this with an effective
clumping factor $\zeta = \langle n^2\rangle / \langle n \rangle$. The
ionization fronts move quickly into voids but slowly into dense
filaments. The ionized mass changes inversely with $\zeta$. Cuts
through the ionized regions generally have ``butterfly" shapes, with
larger $\zeta$ producing more irregular shapes.  Larger scale driving
of the turbulence produces more filamentary emission measure maps. 

We emphasize that our results are based on static density fields, so
they apply only to the early stages of the evolution of an H~{\sc ii}
region when the speed of the ionization fronts remains much larger
than the sound speed of the ionized gas, so that the expansion of the
gas is negligible. (In cases where magnetic field is considered, the neutral
Alfv\'en speed should be used in addition to the ionized sound speed,
but it is rarely higher.) We regard our results as complementary to
the findings of \citet{garcia96}, \citet{williams99}, and
\citet{freyer03}, who study the dynamical evolution of
H~{\sc ii} regions with two-dimensional simulations, but only
with uniform or smoothly varying initial conditions. Future
comprehensive simulations with a full treatment that couples gas
dynamics and radiative transfer, and with realistic initial
conditions, will be necessary to extend these results to later times.

\acknowledgments We thank S. Glover for valuable discussions and a
careful reading of the manuscript, and the anonymous referee for useful
comments and corrections that have helped to improve this paper. The
computations were carried out on the Onyx 2000 computers of the Rose
Center at the American Museum of Natural History (AMNH). This project is
based upon work supported by the National Science Foundation under Grant
No. 0239709, and was partially supported by NSF CAREER grant AST 99-85392,
and NASA grants NAG5-10103 and NAG5-13028.


\begin{thebibliography}{JUNK}

\bibitem[\protect\astroncite{Abel, Norman \& Madau}{1999}]{abel99}
Abel, T, Norman, M. L. \& Madau, P. 1999,
\newblock {ApJ} {523}, 66

\bibitem[\protect\astroncite{Abel}{2000}]{abel00}
Abel, T.  2000,
\newblock {Rev. Mex. Astro. Astrof. Ser. Conf.} {9}, 300 

\bibitem[\protect\astroncite{Axford}{1961}]{axford61}
Axford, W.~I. 1961,
\newblock {Philos. Trans. R. Soc. London Ser. A} {253}, 301

\bibitem[\protect\astroncite{Blitz}{1993}]{blitz93}
Blitz, L. 1993
\newblock {Protostars \& Planets III}, 
\newblock {ed. E. H. Levy \& J. I. Lunine (Tucson: Univ. of Arizona Press)}, 125

\bibitem[\protect\astroncite{Churchwell}{1999}]{churchwell99}
Churchwell E. 1999 In \textit{The Origin of Stars and Planetary Systems},
ed. CJ Lada, ND Kylafis,  Dordrecht, The Netherlands: Kluwer, pp. 515-52.

\bibitem[\protect\astroncite{Churchwell}{2002}]{churchwell02}
Churchwell E. 2002, 
\newblock {ARAA} {40}, 27 

\bibitem[\protect\astroncite{Clarke \& Norman}{1994}]{clarke94}
Clarke, D. \& Norman, M. L. 1994,
\newblock {National Center for Supercomputing Applications Technical Report} 

\bibitem[\protect\astroncite{Crutcher}{1999}]{crutcher99}
Crutcher, R.~M. 1999,
\newblock {ApJ} {520}, 706 

\bibitem[\protect\astroncite{Dyson, Williams \& Redman}{1995}]{dyson95}
Dyson, J. E., Williams, R. J. R. \& Redman, M. P. 1995,
\newblock {MNRAS} {277}, 700

\bibitem[\protect\astroncite{Dyson et al.}{2002}]{dyson02}
Dyson, J. E., Williams, R.~J.~R.,  Hartquist, T.~W. \& Pavlakis, K.~G. 2002, 
\newblock {RevMexAA} {12}, 8

\bibitem[\protect\astroncite{Elmegreen}{1976}]{elmegreen76}
Elmegreen, B.~G. 1976,
\newblock {ApJ} {205}, 405

\bibitem[\protect\astroncite{Elmegreen}{1993}]{elm93} Elmegreen, B. G. 1993, ApJ, 419, L29 

\bibitem[\protect\astroncite{Falle}{2002}]{falle02}
Falle, S.~A.~E.G. 2002
\newblock {ApJ} {577}, L123

\bibitem[\protect\astroncite{Flower}{1969}]{flower69}
Flower,  D.~R. 1969,
\newblock {MNRAS} {146}, 243

\bibitem[\protect\astroncite{Franco et~al.}{2000}]{franco00}
Franco J., Kurtz, S.~E., Garc\'{\i}a-Segura, G. \& Hofner, P. 2000,
\newblock {Ap\&SS} {272}, 169

\bibitem[\protect\astroncite{Freyer, Hensler, \& Yorke}{2003}]{freyer03}
Freyer, T., Hensler, G. \& Yorke, H.~W. 2003,
\newblock {ApJ} {594}, 888 

\bibitem[\protect\astroncite{Garc\'{\i}a-Segura \& Franco}{1996}]{garcia96}
Garc\'{\i}a-Segura, G. \& Franco J. 1996,
\newblock {ApJ} {469}, 171 

\bibitem[\protect\astroncite{Garay \& Lizano}{1999}]{garay99}
Garay G., Lizano S. 1999
\newblock {Publ. Astron. Soc. Pac.} {111}, 1049

\bibitem[\protect\astroncite{Goldsworthy}{1961}]{goldsworthy61}
Goldsworthy, F.~A. 1961,
\newblock {Philos. Trans. R. Soc. London Ser. A} {253}, 277

\bibitem[\protect\astroncite{Hawley \& Stone}{1995}]{hawley95}
Hawley, J.~F. \& Stone, J.~M. 1995 
\newblock {Comput. Phys. Commun.} {89}, 127 

\bibitem[\protect\astroncite{Icke}{1979}]{icke79}
Icke, V. 1979,
\newblock {ApJ} {234}, 615 

\bibitem[\protect\astroncite{Kahn}{1954}]{kahn54}
Kahn, F.~D. 1954,
\newblock {Bull. Astron. Inst. Neth.} {12}, 187

\bibitem[\protect\astroncite{Kirchhoff}{1860}]{kirchhoff60}
Kirchhoff, H. 1860
\newblock {Philos. Mag.} {19}, 193

\bibitem[\protect\astroncite{Kirkpatrick}{1970}]{kirkpatrick70}
Kirkpatrick, R. C. 1970,
\newblock {ApJ} {162}, 33

\bibitem[\protect\astroncite{Kirkpatrick}{1972}]{kirkpatrick72}
Kirkpatrick, R. C. 1972,
\newblock {ApJ} {176}, 381

\bibitem[\protect\astroncite{Klein, Stein \& Kalkofen}{1978}]{klein78}
Klein, R. I., Stein, R. F. \& Kalkofen, W. 1978,
\newblock {ApJ} {220}, 1024 

\bibitem[Klessen, Heitsch, \& Mac Low(2000)]{khm00} Klessen, R. S., Heitsch,
  F., \& Mac Low, M.-M. 2000, ApJ, 535, 887

\bibitem[\protect\astroncite{K\"oppen}{1978}]{koppen78}
K\"oppen, J. 1978,
\newblock {A\&ASS} {35}, 111 

\bibitem[\protect\astroncite{K\"oppen}{1979}]{koppen79}
K\"oppen, J. 1979,
\newblock {A\&A} {80}, 42 

\bibitem[\protect\astroncite{Mac Low}{1999}]{maclow99}
Mac Low, M-M. 1999,
\newblock {ApJ} {524}, 169 

\bibitem[ Mac Low \& Klessen(2004)]{mk04} Mac Low, M.-M., \& Klessen,
  R. S. 2004, Rev.\ Mod.\ Phys., 76, 125 

\bibitem[\protect\astroncite{Marsh}{1970}]{marsh70}
Marsh, M.~C. 1970,
\newblock {MNRAS} {147}, 95 

\bibitem[Ostriker, Stone, \& Gammie(2001)]{osg01} Ostriker, E. C., Stone,
  J. M., \& Gammie, C. F. 2001, ApJ 546, 9800 

\bibitem[\protect\astroncite{Panagia}{1973}]{panagia73}
Panagia, N. 1973,
\newblock {AJ} {78}, 929 

\bibitem[\protect\astroncite{Pequignot, Stasinska \& Aldrovandi}{1978}]{pequignot78}
Pequignot, D., Stasinska, G. \& Aldrovandi, S. M. V. 1978
\newblock {A\&A} {63}, 313

\bibitem[\protect\astroncite{Redman et~al}{1998}]{redman98}
Redman, M. P., Williams, R. J. R., Dyson, J. E.; Hartquist, T. W. \&
Fernandez, B. R. 1998,
\newblock {A\&A} {331}, 1099

\bibitem[\protect\astroncite{Sandford}{1973}]{sandford73}
Sandford, M. T., II 1973,
\newblock {ApJ} {183}, 555

\bibitem[\protect\astroncite{Sandford, Whitaker \& Klein}{1982}]{sandford82}
Sandford, M. T., II, Whitaker, R. W. \& Klein, R. I. 1982,
\newblock {ApJ} {260}, 183

\bibitem[\protect\astroncite{Spitzer}{1978}]{spitzer78}
Spitzer, L. Jr. 1978,
\newblock {Physical Processes in the Interstellar Medium (New York: John Wiley
\& Sons)}

\bibitem[\protect\astroncite{Stone \& Norman}{1992a}]{stone92a}
Stone, J. M. \& Norman, M. L. 1992,
\newblock {ApJS} {80}, 753

\bibitem[\protect\astroncite{Stone \& Norman}{1992b}]{stone92b}
Stone, J. M. \& Norman, M. L. 1992,
\newblock {ApJS} {80}, 791

\bibitem[\protect\astroncite{Stromgren}{1939}]{stromgren39}
Str\"omgren, B. 1939,
\newblock {ApJ} {89}, 526

\bibitem[\protect\astroncite{Tenorio-Tagle}{1976}]{tenorio-tagle76}
Tenorio-Tagle, G. 1976,
\newblock {A\&A} {53}, 411 

\bibitem[\protect\astroncite{Tenorio-Tagle}{1979}]{tenorio-tagle79}
Tenorio-Tagle, G. 1979,
\newblock {A\&A} {71}, 59 

\bibitem[\protect\astroncite{Tenorio-Tagle \& Bedijn}{1981}]{tenorio-tagle81}
Tenorio-Tagle, G. \& Bedijn, P.~J. 1981,
\newblock {A\&A} {99}, 305 

\bibitem[\protect\astroncite{van Leer}{1977}]{vanleer77}
van Leer, B. 1977, 
\newblock {J. Comput. Phys.} {23}, 276 

\bibitem[\protect\astroncite{Williams}{1999}]{williams99}
Williams, R. J. R. 1999,
\newblock {MNRAS} {310}, 789

\bibitem[\protect\astroncite{Williams, Dyson \& Hartquist}{2000}]{williams00}
Williams, R. J. R., Dyson, J. E. \& Hartquist, T. W. 2000,
\newblock {MNRAS} {314}, 315 

\bibitem[\protect\astroncite{Wood \& Churchwell}{1989}]{wood89}
Wood, D. O. S. \& Churchwell E. 1989,
\newblock {ApJS} {69}, 831

\bibitem[\protect\astroncite{Yang et al.}{1996}]{yang96}
Yang, H., Chu, Y.; Skillman, E.~D. \& Terlevich, R. 1996
\newblock {ApJ} {112}, 146

\bibitem[\protect\astroncite{Yorke}{1986}]{yorke86}
Yorke, H.~W. 1986,
\newblock {ARAA} {24}, 49 

\end{thebibliography}
\end{document}